\documentclass[sigconf]{acmart}
\AtBeginDocument{%
  }

\begin{document}

\copyrightyear{2026}
\acmYear{2026}
\setcopyright{cc}
\setcctype{by}
\acmConference[ISLPED '26]{ACM/IEEE International Symposium on Low Power Electronics and Design}{August 05--07, 2026}{Evanston, IL, USA}
\acmBooktitle{ACM/IEEE International Symposium on Low Power Electronics and Design (ISLPED '26), August 05--07, 2026, Evanston, IL, USA}
\acmDOI{10.1145/3816440.3818598}
\acmISBN{979-8-4007-2748-1/2026/08}

\title[Unified Complex-Valued Acceleration for Matrix-Based HDC]%
      {Vector-Space Projection and Unified Complex-Valued Acceleration for Scaling Matrix-Based Brain-Inspired Representations}


\author{William Youngwoo Chung}
\email{chungwy1@uci.edu}
\affiliation{%
  \institution{Information and Computer Sciences}
  \institution{University of California Irvine}
  \city{Irvine}
  \state{California}
  \country{USA}
}

\author{Hyunwoo Oh}
\email{hyunwooo@uci.edu}
\affiliation{%
  \institution{Information and Computer Sciences}
  \institution{University of California Irvine}
  \city{Irvine}
  \state{California}
  \country{USA}
}

\author{Calvin Yeung}
\email{chyeung2@uci.edu}
\affiliation{%
  \institution{Information and Computer Sciences}
  \institution{University of California Irvine}
  \city{Irvine}
  \state{California}
  \country{USA}
}

\author{Hansen Jin Lillemark}
\email{hlillemark@ucsd.edu}
\affiliation{%
  \institution{Computer Science and Engineering}
  \institution{University of California, San Diego}
  \city{La Jolla}
  \state{California}
  \country{USA}
}

\author{Hamza Errahmouni Barkam}
\email{herrahmo@uci.edu}
\affiliation{%
  \institution{Information and Computer Sciences}
  \institution{University of California Irvine}
  \city{Irvine}
  \state{California}
  \country{USA}
}

\author{Mohsen Imani}
\email{m.imani@uci.edu }
\affiliation{%
  \institution{Information and Computer Sciences}
  \institution{University of California Irvine}
  \city{Irvine}
  \state{California}
  \country{USA}
}

\renewcommand{\shortauthors}{Chung et al.}

\begin{abstract}
Symbolic reasoning at scale requires representations that are both expressive and hardware-efficient. Hyperdimensional Computing (HDC) provides a promising brain-inspired framework, encoding concepts as high-dimensional vectors composed through efficient algebraic operations. However, the widely used Fourier Holographic Reduced Representation (FHRR) relies on commutative binding, which cannot encode order or directionality. Prior FPGA/ASIC accelerators recover this through costly permutation logic, incurring irregular memory access and high control overhead. Generalized Holographic Reduced Representations (GHRR) instead redefine binding as non-commutative matrix multiplication, naturally encoding sequences, graphs, and hierarchies without permutations—but at the cost of turning every element into a matrix, introducing compute and storage overhead that limits scalability. In this work, we propose a flattening methodology that preserves GHRR's matrix-based encoding while executing training and inference directly in vector space, functionally equivalent to FHRR inference yet free of permutation logic. We further design a custom 28 nm ASIC that fuses binding and similarity into a unified complex-valued datapath, with dual-DMA streaming and runtime normalization for accurate inference. Against a PyTorch baseline on an NVIDIA RTX 4090 GPU, the prototype delivers 1.36×–1.56× higher throughput and 16.2×–18.6× better energy efficiency.

\end{abstract}



\maketitle

\section{Introduction} \label{sec:intro}
Designing architectures for symbolic reasoning requires representations to be expressive enough to capture complex structures, while also being efficient enough to meet the power and latency budgets of modern accelerators. Machine learning models have seen excellent results across various domains but often entangle objects and attributes into dense activations, limiting their transparency and compositional reasoning ability. For deployment in neuromorphic and embedded systems, an alternative framework is needed, one that preserves symbolic structure while working within hardware limitations.

Hyperdimensional Computing (HDC), or Vector Symbolic Architectures (VSA), offers such a framework by encoding concepts as high-dimensional vectors and manipulating them through algebraic operations such as binding, unbinding, and superposition~\cite{kanerva2009hyperdimensional}. These operations are inherently parallel and noise-tolerant, making them attractive for hardware acceleration. Yet the most widely used implementation, Fourier Holographic Reduced Representation (FHRR)\cite{Plate1995HRR}, relies on commutative based binding, which cannot capture order and directionality. Sequences and graphs must therefore be encoded through additional permutation logic. On hardware, this requirement results in costly, irregular memory access, which can undermine the efficiency of prior FPGA and ASIC accelerators~\cite{imani2019fach, chen2023hypergraf}.

Generalized Holographic Reduced Representation (GHRR) \cite{yeung2024ghrr} overcomes this bottleneck by redefining binding as matrix multiplication in $\mathbb{C}^{d\times d}$. Unlike commutative element-wise products, matrix multiplication is non-commutative such that embedding structure and order can be directly done in the representation space. Beyond its representational benefits, GHRR’s core operations align naturally with GEMM-like operations, a primitive already optimized across accelerator architectures. By collapsing compositional reasoning into GEMM-style datapaths and non-commutative binding, GHRR eliminates the need for permutation-specific hardware and enables symbolic reasoning to scale on existing high-throughput pipelines.

This work introduces a permutation-free accelerator design that natively supports GHRR. Rather than positional embeddings or permutations, we reframe the encoding symbolic reasoning as a series of matrix multiplications, allowing binding, unbinding, and similarity to execute directly on a unified complex-valued datapath. Our 28\,nm ASIC prototype integrates runtime normalization, interleaved real–imaginary buffers, and dual DMA streaming to sustain high throughput, achieving 1.54\, TFLOPS peak performance within 4.77\,mm² and 4.22\,W. Compared to a PyTorch baseline on an NVIDIA RTX 4090 GPU, the design delivers $1.36{\text{–}}1.56\times$ higher throughput and $16.2{\text{–}}18.6\times$ energy efficiency.

By rethinking GHRR through a hardware lens, we demonstrate that the path to efficient symbolic reasoning can be achieved through additional parallelism or memory optimization, and eliminating permutation altogether. This permutation-free methodology simplifies accelerator design, scales predictably with dimensionality and parallelism, and provides a practical foundation for embedding brain-inspired reasoning into next-generation machine learning hardware.

In this work, we present a unified algorithm--architecture framework for scalable, permutation-free GHRR acceleration. First, we show that GHRR similarity can be computed exactly in a flattened complex vector space, enabling efficient training and inference without explicit matrix-valued computation. Second, we design a streaming ASIC accelerator that maps binding, unbinding, and similarity onto a unified complex-valued datapath. Finally, we demonstrate that the proposed design achieves FHRR-level inference efficiency while preserving GHRR's non-commutative representational advantages.

\section{Background \& Related Works} 
\label{sec:background}

\subsection{Hyperdimensional Computing for Structured Data}
Hyperdimensional Computing (HDC) represents symbolic entities as high-dimensional vectors and manipulates them through algebraic operations such as \emph{bundling} (superposition) and \emph{binding} (association)~\cite{stock2024hyperdimensional, kanerva2009hyperdimensional, zou2022memory, zou2022eventhd, zou2021spiking, imani2023hierarchical, poduval2022graphd, zou2021scalable}. These operations are massively parallel, noise-tolerant, and hardware-friendly, enabling applications in domains such as edge intelligence, robotics, and reinforcement learning~\cite{neubert2019introduction, hd_20, snyder2025generalizable, ni2024efficient, bhattacharjya2026hyperdoa}.

Holographic Reduced Representations (HRR)~\cite{Plate1995HRR} are one of the most established forms of HDC, defining binding as circular convolution, bundling as addition, and similarity as dot product between high-dimensional vectors. The Fourier variant (FHRR)~\cite{plate2003HRR} replaces convolution with element-wise complex multiplication, which allows exact unbinding through conjugation and further simplifies hardware realization. Both HRR and FHRR preserve distributed and robust encodings and form the foundation of many neurosymbolic HDC-based frameworks. However, because element-wise multiplication is commutative, these encodings cannot naturally capture order or directionality. To represent sequences or directed graphs, additional permutation operations are typically employed~\cite{frady2018theory}, which complicate interpretability, hardware mapping, and information retrieval \cite{kleyko2022survey, schlegel2022comparison, kleyko2023survey}. 

More recently, graph-centric HDC methods have pursued graph-based symbolic reasoning. 
MoleHD~\cite{ma2021molehd} embeds molecular graphs into hypervectors at low energy cost, GrapHD~\cite{poduval2022graphd} encodes entire graph structures as single hypervectors, and HDReason~\cite{chen2024hdreason} and RelHD~\cite{kang2022relhd} incorporate relational and multi-hop information; recent work has also examined reasoning quality and bias in graph HDC~\cite{liu2025hypervectors, liu2025mitigating}. Despite their success, these methods remain reliant on commutative binding operators and thus fall back on permutations or handcrafted role vectors to encode structural information. This reliance introduces additional hardware overheads and limits compositional reasoning efficiency.

\subsection{Generalized Holographic Reduced Representations}
To address the problem of HDC encoding not naturally capturing order~\cite{gallant2013representing, gosmann2019vector, yeung2024ghrr}, various recent solutions have been proposed. GHRR address the commutativity limitation by extending FHRR into the matrix domain. Instead of unitary complex scalars $e^{i\theta} \in U(1)$, GHRR employs unitary matrices $a_j \in U(m)$, yielding hypervectors represented as tensors $H \in \mathbb{C}^{D \times m \times m}$, where m is defined as the complexity. Binding is defined as matrix multiplication and bundling as addition, with similarity computed via normalized Frobenius inner products. When $m=1$, GHRR reduces exactly to FHRR, making it a natural generalization. This formulation retains the robustness and distributed properties of HDC while enabling non-commutative binding and richer representational capacity. It also broadens the kernel approximation view of FHRR, which is closely related to Random Fourier Features (RFF)~\cite{rahimi2008random}. Through Bochner’s theorem, FHRR encodings approximate shift-invariant kernels such as the RBF kernel when Gaussian sampling is used~\cite{thomas2021theoretical}. GHRR inherits this kernel interpretation but extends it by supporting matrix-valued encodings that can realize a wider family of kernels.  

Despite its theoretical advantages, empirical demonstrations of GHRR have so far been limited to small-scale tests of non-commutativity and decoding accuracy. No prior work has systematically benchmarked its performance on large-scale workloads, nor have its hardware implications been studied. Our work fills this gap by presenting the first hardware software acceleration co-design of GHRR by introducing a flattening technique alongside a custom accelerator design that maintains hardware efficiency while exploiting GHRR’s non-commutative algebra.

\begin{figure*}[h]
    \centering
    \vspace{-2mm}
    \includegraphics[width=0.8\textwidth]{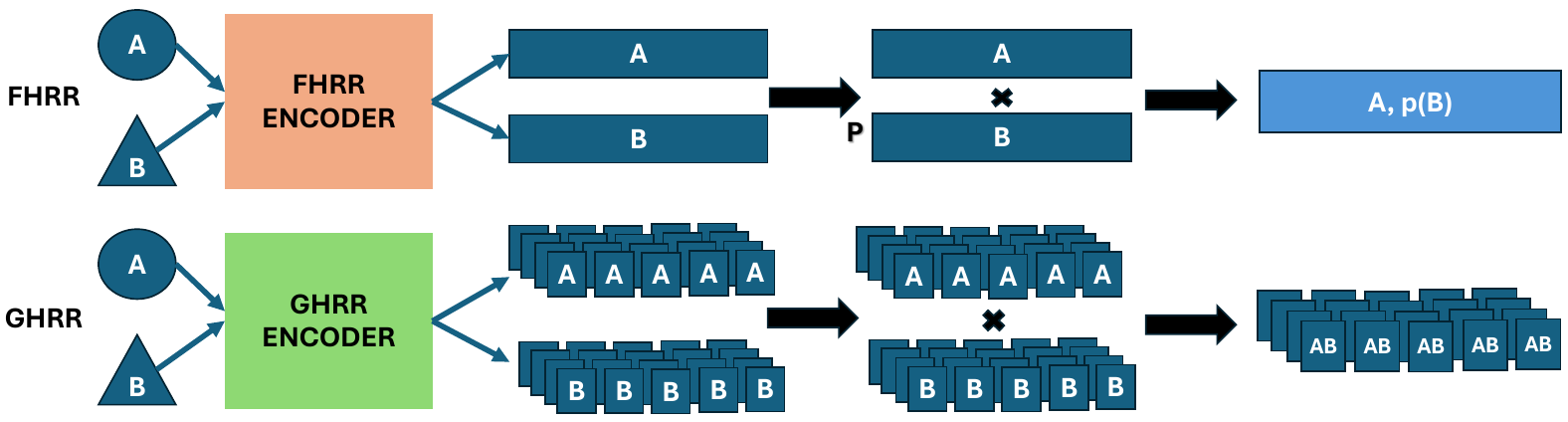}
    \vspace{-2mm}
    \caption{FHRR relies on permutations to encode structure while GHRR can encode direction through non-commutative binding. As a result, GHRR's representations are better fit for encoding complex structures whereas FHRR requires hand-engineered permutation operations.}
    \label{fig:GHRR}
    \vspace{-2mm}
\end{figure*}

\subsection{Hardware Acceleration for HDC}
Research on hardware acceleration of HDC has spanned FPGA, PIM, and ASIC platforms~\cite{kleyko2022vector, karunaratne2020memory}. Early efforts targeted lightweight, low-power classification tasks by simplifying HDC algebra into binary or real-valued operations for embedded and edge devices~\cite{imani2019fach, ni2022algorithm, kleyko2012dependable, kleyko2019integer, oh2026torr}. These designs leveraged the inherent parallelism of bundling and similarity search but offered limited support for structured data. A complementary line of work instead targets the storage cost of HDC inference itself, compressing class representations under extreme memory budgets through decomposed or logarithmically reduced codebooks~\cite{yun2025decohd, yun2025loghd}, an axis of scalability orthogonal to the binding operator targeted in this work.

More recent work has focused on graph reasoning with HDC. HyperGRAF~\cite{chen2023hypergraf} introduced a domain-specific accelerator for graph-based HDC, demonstrating significant efficiency gains but relying on commutative binding and additional control for role encodings. RelHD~\cite{kang2022relhd} and HDReason~\cite{chen2024hdreason} propose memory-centric designs that embed role binding directly into FeFET arrays or systolic fabrics, achieving scalability for undirected graphs. However, these architectures remain tied to commutative operators, which forces the use of handcrafted role vectors or positional permutations to represent directionality. As a result, this reliance inflates area, latency, and design complexity. Hardware accelerators for HDC frequently optimize only the bundling stages (e.g., Hadamard transforms), while overlooking the core binding operator that governs structure-preserving composition.

In contrast, our approach eliminates permutation entirely by aligning GHRR's non-commutative binding with dense matrix operations and flattening its matrix space into a vector space better supported by HDC hardware. This matters because permutations, though trivial in software, are a major hardware bottleneck: a permutation moves all elements of a hypervector across the memory hierarchy with negligible computation, yielding low arithmetic intensity. On GPUs this disrupts access and tensor-core utilization; on ASICs and FPGAs it forces general-purpose shuffle or crossbar networks~\cite{chen2023hypergraf} that consume routing resources and add latency. This ties directly to the memory wall, where DRAM/HBM traffic dominates accelerator energy and execution time~\cite{chung2025robust}. By replacing permutations with non-commutative binding, our design reduces off-chip traffic and yields a compute–memory balance better suited to modern accelerators, enabling hardware-friendly symbolic reasoning that scales to sequential and relational tasks while retaining the throughput expected of HDC accelerators.

\subsection{GHRR Mathematics}

A GHRR hypervector is composed of $D$ unitary matrix blocks of size $m \times m$:
\begin{equation}
\mathbf{H}
= [\mathbf{A}^{(1)},\ldots,\mathbf{A}^{(D)}]^\top
\in \mathbb{C}^{D\times m\times m},
\end{equation}
where each $\mathbf{A}^{(j)} \in \mathrm{U}(m)$ is drawn from a distribution $p$ over unitary matrices.  
Here, $D$ is the dimensionality of the hypervector itself whereas $m$ is defined as the complexity and size of the matrix.

Bundling (superposition) in GHRR is defined as element-wise addition:
\begin{equation}
\mathbf{H}_1 + \mathbf{H}_2
= [\mathbf{A}^{(j)} + \mathbf{B}^{(j)}]_{j=1}^D,
\end{equation}
whereas binding (association) is defined as element-wise matrix multiplication:
\begin{equation}
\mathbf{H}_1 * \mathbf{H}_2
= [\mathbf{A}^{(j)} \mathbf{B}^{(j)}]_{j=1}^D,
\end{equation}
which is non-commutative and therefore captures directionality between associations. Similarity (retrieval) uses the Frobenius inner product as the similarity metric:
\begin{equation}
\delta(\mathbf{H}_1,\mathbf{H}_2)
= \frac{1}{mD}
    \Re\!\left[
    \sum_{j=1}^D
    \mathrm{tr}\!\big(
      \mathbf{A}^{(j)}
      (\mathbf{B}^{(j)})^\dagger
    \big)
    \right],
\label{eq:ghrr-sim}
\end{equation}
where $\Re$ refers to taking the real component and $\mathrm{tr}$ is the trace of the resulting matrix averaged across $D$.

\subsection{Flattening Equivalence}
Matrix-valued representations increase expressivity but introduce computational overhead during training and inference. The key observation enabling our hardware design is that GHRR similarity depends only on the \emph{Frobenius inner products} of the blocks, which admit an exact vector form:
\begin{equation}
\langle \mathbf{X},\mathbf{Y}\rangle_F
= \mathrm{tr}(\mathbf{X}^\dagger\mathbf{Y})
= \mathrm{vec}(\mathbf{X})^\dagger \mathrm{vec}(\mathbf{Y}).
\end{equation}

Therefore, for any $\mathbf{A},\mathbf{B}\in\mathbb{C}^{m\times m}$,
\begin{align}
\mathrm{tr}(\mathbf{A}\mathbf{B}^\dagger)
\;=\; \mathrm{tr}\big((\mathbf{B}\mathbf{A}^\dagger)^\dagger\big)
\;=\; \mathrm{vec}(\mathbf{B})^\dagger \mathrm{vec}(\mathbf{A}).
\label{eq:swap-vec}
\end{align}
Applying \eqref{eq:swap-vec} blockwise to $\mathbf{H}_1=[\mathbf{A}^{(j)}]_{j=1}^D$ and $\mathbf{H}_2=[\mathbf{B}^{(j)}]_{j=1}^D$ gives the following:
\begin{align}
\sum_{j=1}^D \mathrm{tr}\!\big(\mathbf{A}^{(j)}(\mathbf{B}^{(j)})^\dagger\big)
&= \sum_{j=1}^D \mathrm{vec}(\mathbf{B}^{(j)})^\dagger
   \,\mathrm{vec}(\mathbf{A}^{(j)}) \notag \\
&= \big(\mathrm{flat}(\mathbf{H}_2)\big)^\dagger
   \,\mathrm{flat}(\mathbf{H}_1),
\end{align}
where
\begin{equation}
\mathrm{flat}(\mathbf{H})
\in \mathbb{C}^{Dm^2}
\end{equation}
concatenates the vectorized blocks.
Substituting into \eqref{eq:ghrr-sim} gives:
\begin{equation}
\delta(\mathbf{H}_1,\mathbf{H}_2)
= \frac{1}{mD}\,
\Re\!\left[
\big(\mathrm{flat}(\mathbf{H}_2)\big)^\dagger
\mathrm{flat}(\mathbf{H}_1)
\right].
\label{eq:flat-eq}
\end{equation}
Since the two inner products are complex conjugates,
\begin{align}
\label{eq:ghrr-frob}
\Big(\mathrm{flat}(\mathbf{H}_2)\Big)^\dagger \mathrm{flat}(\mathbf{H}_1)
\;=\; \overline{\Big(\mathrm{flat}(\mathbf{H}_1)\Big)^\dagger \mathrm{flat}(\mathbf{H}_2)}.
\end{align}
Taking real part of \eqref{eq:ghrr-frob} yields
\begin{align}
\delta(\mathbf{H}_1,\mathbf{H}_2)
&= \frac{1}{mD}\,\Re\!\left[\sum_{j=1}^{D}\mathrm{tr}\!\big(\mathbf{A}^{(j)}(\mathbf{B}^{(j)})^\dagger\big)\right] \\
&= \frac{1}{mD}\,\Re\!\left[\Big(\mathrm{flat} (\mathbf{H}_1)\Big)^\dagger \mathrm{flat}(\mathbf{H}_2)\right].
\label{eq:flat-sim}
\end{align}
Therefore, the normalized similarity in the matrix space equals a normalized real part of an inner product in the flattened vector space. This justifies training and inference directly on $\mathrm{flat}(\mathbf{H})\in\mathbb{C}^{Dm^2}$ with standard high-throughput dot-product engines, without losing the GHRR semantics.

\subsection{Utilizing GHRR Encoding}

Now that we have established that GHRR can be flattened for inference, we will now explain GHRR encoding works. Given a data point $\mathbf{x}\in\mathbb{R}^n$, we construct its GHRR representation blockwise as
\begin{align}
\phi(\mathbf{x})^{(j)} \;=\; \mathbf{Q}^{(j)} \mathbf{\Lambda}^{(j)}(\mathbf{x}), 
\quad j=1,\ldots,D,
\end{align}
where each $\mathbf{Q}^{(j)} \in \mathrm{U}(m)$ is a unitary matrix (either fixed or varying across $j$). The diagonal modulation is defined by
\begin{align}
\mathbf{\Lambda}^{(j)}(\mathbf{x})
\;=\; \exp\!\Big(i\,\mathrm{diag}(\theta_{1j}^\top \mathbf{x}, \ldots, \theta_{mj}^\top \mathbf{x})\Big),
\end{align}
with vectors, $\theta_{ij}$, drawn from a distribution $p$. For two inputs $\mathbf{x},\mathbf{y}\in\mathbb{R}^n$, the similarity
\(\delta\!\big(\phi(\mathbf{x}),\phi(\mathbf{y})\big)\) yields an approximation to a kernel $K(\mathbf{x},\mathbf{y})$.

After the GHRR encoding, we use $\mathbf{h}=\mathrm{flat}\big(\phi(\mathbf{x})\big)\in\mathbb{C}^{Dm^2}$ in any standard classifier or HDC routine. Bundling, updates, and similarity operate identically in the flattened domain because of \eqref{eq:flat-sim}. This results in an \emph{effective dimension} of $D_E=Dm^2$, preserving GHRR’s non-commutative binding association while enabling learning on an efficient vector space for hardware acceleration.

\section{Hardware Acceleration}
Now that GHRR can be mapped into a vector space for inference, we co-design a hardware accelerator tailored to its core operations: binding-based encoding and similarity-based inference, making GHRR viable for deployment in low-latency or resource-constrained environments. Unlike traditional HDC systems that rely on commutative binding and permutation logic, our ASIC design natively supports GHRR’s matrix-based, non-commutative algebra while maintaining end-to-end differentiability and throughput scalability.

\begin{figure}[h]
    \centering
    \includegraphics[width=1\linewidth]{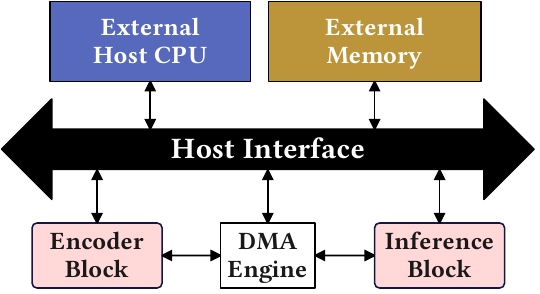}
    \caption{Top-level architecture of the GHRR accelerator. Host interface coordinates encoder and inference blocks via DMA streaming from external memory.}
    \label{fig:ghrr_top_arch}
\end{figure}

The overall system architecture is shown in Fig.~\ref{fig:ghrr_top_arch}. The accelerator consists of three main subsystems: the \textit{Encoder Block}, which transforms input features into hyperdimensional GHRR representations; the \textit{Inference Block}, which evaluates normalized similarity against a learned codebook; and a control/data interface managed by a lightweight \textit{Host Interface}. Communication with off-chip memory is handled by a dual-channel \textit{DMA Engine}, which streams both raw inputs and codebook vectors from external memory into the on-chip buffers.

\subsection{Encoder Block}
\begin{figure}[h]
    \centering
    \includegraphics[width=\linewidth]{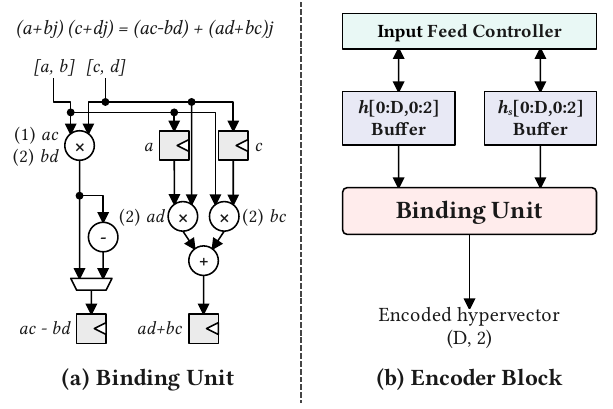}
    \vspace{-2mm}
    \caption{Encoder acceleration block}
    \label{fig:ghrr_encoder}
    \vspace{-2mm}
\end{figure}

The encoder module translates input data into GHRR hypervectors using a complex-valued transformation pipeline. As shown in Fig.~\ref{fig:ghrr_encoder}, the core of this pipeline is a \textit{Binding Unit} that multiplies pairs of complex-valued matrices element-wise using the standard rule for complex multiplication:
\[
(a + bj)(c + dj) = (ac - bd) + (ad + bc)j
\]
This multiplication is implemented via parallel fused-multiply-add (FMA) units, with real and imaginary parts stored in dual buffers of shape $(D, 2)$, where $D$ is the hyperdimension. Each input vector is projected through a bank of fixed or randomized unitary transformations $Q(j)$, followed by an element-wise complex exponential controlled by $\Lambda^{(j)}(x)$. These steps are pipelined for latency hiding and throughput maximization, with the resulting encoding streamed directly into downstream inference buffers.

\subsection{Inference Block}
\begin{figure}[h]
    \centering
    \includegraphics[width=\linewidth]{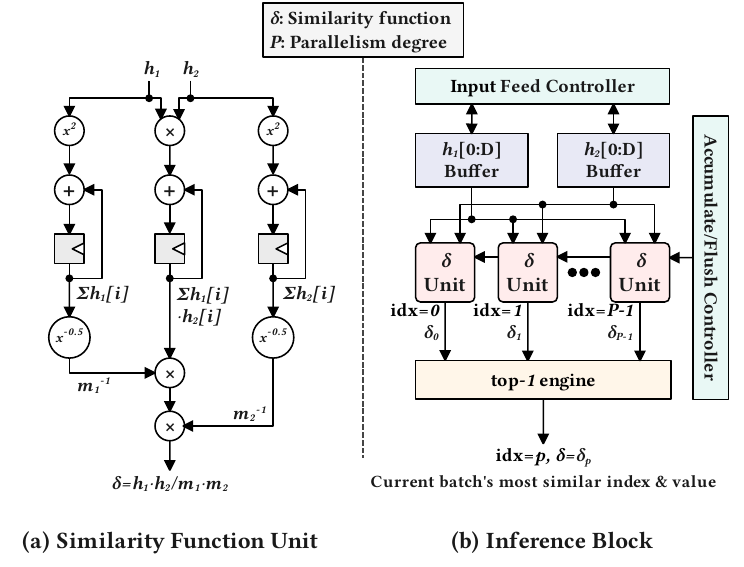}
    \vspace{-5mm}
    \caption{Inference acceleration block}
    \label{fig:ghrr_inference}
    \vspace{-3mm}
\end{figure}

Inference proceeds by computing the similarity between the current encoding and a fixed hyperdimensional codebook. In hardware, this is realized by a fully parallel \textit{Similarity Function Unit} (Fig.~\ref{fig:ghrr_inference}(a)) that compares $P$ candidate vectors in parallel. Each comparison computes a normalized dot product:
\[
\delta(h_1, h_2) = \frac{h_1^{\text{real}} \cdot h_2^{\text{real}}}{\|h_1^{\text{real}}\| \cdot \|h_2^{\text{real}}\|}
\]
Here, $h_1^{\text{real}}$ and $h_2^{\text{real}}$ are interleaved real-valued encodings of the original complex hypervectors. The system avoids complex conjugation for efficiency and instead directly evaluates the real-valued similarity approximation. This approximation maps cleanly to vector dot product units and eliminates the need for complex arithmetic.

The norm of each input vector is computed at runtime using accumulation ($\Sigma h[i]^2$), and their multiplicative inverses $m_1^{-1}$ and $m_2^{-1}$ are generated in hardware. These values scale the final dot product through two multiplication stages. The result of each $\delta$ unit is forwarded to a lightweight \textit{top-1 engine}, which identifies the index of the most similar codebook vector across all $P$ candidates in the current batch.

This runtime normalization is essential: unlike Hermitian inner products, the dropped conjugation means that $\|h\|$ is not fixed to the dimensionality $D$, and may vary significantly based on the input distribution. Without normalization, similarity scores would be biased by magnitude differences, impairing robustness and accuracy.

To support high-throughput inference, the reciprocal square root $\|h\|^{-1}$ is computed on-chip using a pipelined logic based on the Newton-Raphson method. This value is then refined across multiple stages of the pipeline to produce the final $\|h\|^{-1}$, which is used to scale the accumulated dot product.

The inference pipeline overlaps input loading, dot product evaluation, norm accumulation, reciprocal square root computation, and top-$1$ selection with minimal control overhead. This architecture maintains high utilization even as batch size and dimensionality increase.

\subsection{Throughput and Parallelism}
Both the encoder and inference blocks are designed to support data-level parallelism. The pipeline architecture overlaps data fetch, binding, and accumulation phases, while a streaming input controller ensures high utilization across all compute units. Our accelerator supports configurable parallelism degree $P$, enabling adaptation to varying batch sizes and real-time throughput requirements without sacrificing accuracy.

\subsection{Advantages Over Permutation-Based Designs}
By eliminating the need for positional permutations or synthetic encodings, our hardware design simplifies memory access patterns and reduces logic area. All compositional reasoning steps, binding, unbinding, and comparison, collapse into GEMM-like operations with minimal control overhead. As a result, GHRR’s algebra maps cleanly onto both spatial (e.g., systolic) and temporal (e.g., pipelined) architectures. The matrix-based encoding allows us to recover order and hierarchy directly from the algebra, avoiding the combinatorial explosion of permutation states typical in traditional FHRR accelerators.

\section{Results}
\subsection{Inference Experiments}

To validate our flattening method, we benchmark FHRR, GHRR (Flatten) and GHRR (Matrix) on a synthetic sequence classification dataset with samples of sequence of length = 5 and 10 possible classes. The goal is to measure the inference cost of a given naive GHRR implementation when compared to FHRR and GHRR (Flatten). We run every model on a NVIDIA RTX 3060Ti GPU for 50 trials and report the average inference times.

\begin{figure}[h]
    \centering    \includegraphics[width=0.8\linewidth]{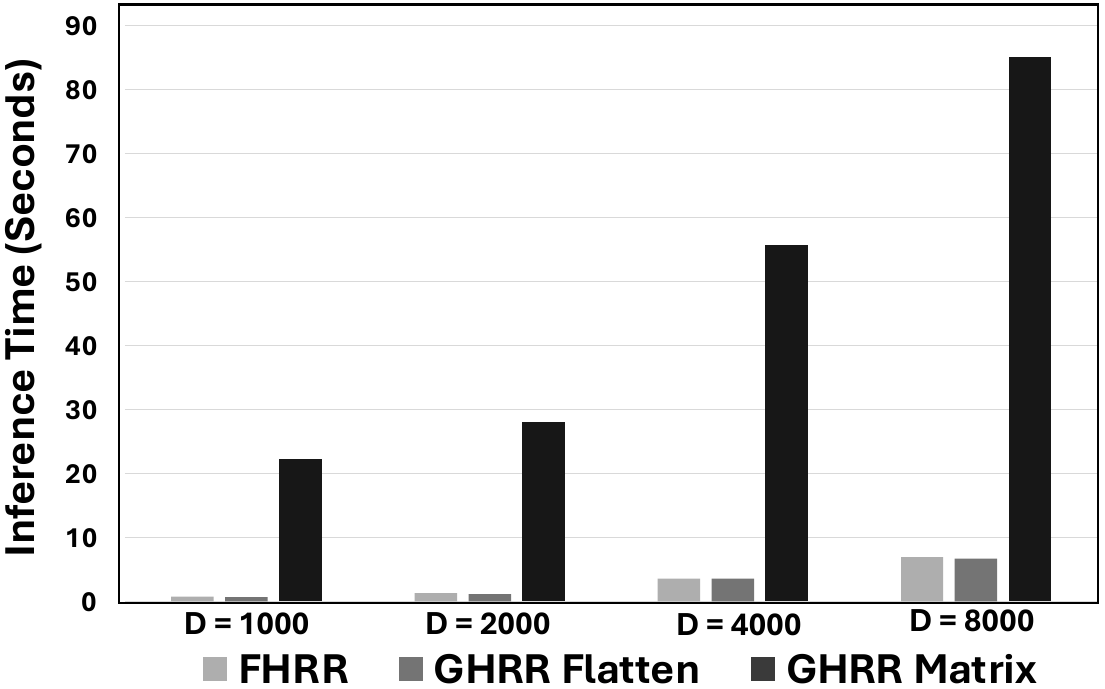}
    \vspace{-1mm}
    \caption{Inference cost of FHRR, GHRR Flatten, and GHRR Matrix on a synthetic sequence based dataset.}
    \label{fig:flatten}
    \vspace{-1mm}
\end{figure}

We compare the inference speeds between all three implementations to gauge how efficient flattening after encoding is relative to FHRR and a normal GHRR implementation as well as the effect of hyperdimension $D$ on the inference time in Figure~\ref{fig:flatten}. In our experiments we note that GHRR (Matrix) takes exponentially longer during inference whereas GHRR (Flatten) remains equivalent speeds as FHRR. Across all dimensions, GHRR (Flatten) remains within $0.95\times$ - $1.05\times$ the inference time as FHRR. On the other hand, GHRR's (Matrix) inference times relative to FHRR is $38.2\times$ at $D=1000$, $16.8\times$ at $D=2000$, $15.5\times$ at $D=4000$, and $12.3\times$ at $D=8000$, showing an extreme cost of keeping GHRR representations as matrices. However, we do note that the gap between the inference times does decrease as $D$ increases. Figure~\ref{fig:flatten} confirms our motivation to accelerate GHRR through vector space flattening for a scalable non-commutative-based HDC learning frameworks.

\subsection{Hardware Acceleration}
In this section, we implement the custom GHRR accelerator as a register-transfer level (RTL) design in Chisel, enabling modular and flexible architectural configurations. Functional verification is conducted using Verilator 5.009, and the design is synthesized using Synopsys Design Compiler targeting the 28\,nm CMOS process, meeting the timing constraints at 1.0\,GHz.

\subsubsection{Performance and Efficiency}
To assess the practicality of our architecture on reasonable HDC workloads, we synthesize the full GHRR pipeline in Fig. \ref{fig:ghrr_top_arch} and evaluate area, power, and throughput. Experiments use a reasonable default hardware configuration of hyperdimension $D=8000$, complexity $M=2$, and parallelism $P=8$, aligned with the benchmark workloads in this section. These settings reflect the target deployment regime of the accelerator and enable direct comparison against state-of-the-art GPU implementation.

Table~\ref{tab:hw_spec} summarizes the resulting specifications. The complete system occupies 4.77\,mm$^2$ of silicon area, operates at 1.00\,GHz, and achieves a peak throughput of 1.54\,TFLOPS, with a maximum power draw of 4.22\,W.

\begin{table}[h]
\centering
\caption{Hardware Specifications}
\vspace{-2mm}
\resizebox{\linewidth}{!}{%
\begin{tabular}{c|c|c|c|c}
\toprule
\textbf{Parameter}         & \textbf{Total Area} & \textbf{Max Power} & \textbf{Frequency} & \textbf{Peak Throughput} \\ \midrule \midrule
\textbf{Value}             & 4.77 mm$^2$         & 4.22 W             & 1.00 GHz           & 1.54 TFLOPS              \\ 
\bottomrule
\end{tabular}%
}
\label{tab:hw_spec}
\end{table}

\begin{figure}[h]
    \centering    \includegraphics[width=\linewidth]{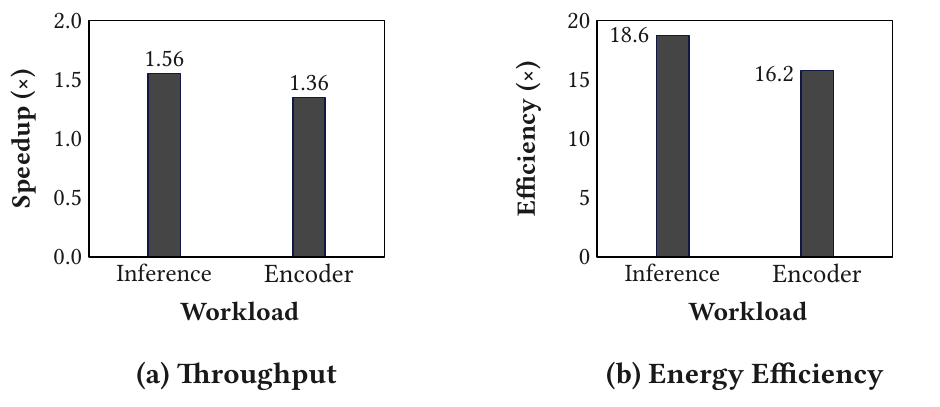}
    \vspace{-5mm}
    \caption{Comparison of throughput and energy efficiency against an NVIDIA RTX 4090 baseline.}
    \label{fig:hw_results}
    \vspace{-3mm}
\end{figure}

We benchmark our ASIC design against an optimized software baseline running on an NVIDIA RTX 4090 GPU. As shown in Fig.~\ref{fig:hw_results}, the accelerator achieves up to $1.56\times$ higher throughput for encoding and $1.36\times$ for inference. The energy efficiency gains are even more substantial, with $18.6\times$ and $16.2\times$ reductions in energy per operation, respectively.

These improvements arise from architectural co-design that fuses complex-valued HDC primitives into a single streamlined datapath, eliminating permutation-heavy shuffles common in prior work. With on-the-fly normalization and a fully pipelined binding–similarity flow, the design sustains high utilization across \(D\) and batch sizes, translating algorithmic regularity directly into throughput and energy gains.

\subsubsection{Large Dimensionality Ablation} 
To thoroughly test the scalability of our proposed accelerator, we additionally benchmark our ASIC across hypervector dimensions 
\(D \in \{16\mathrm{K}, 32\mathrm{K}, 64\mathrm{K}, 128\mathrm{K}, 256\mathrm{K}\}\) as a definitive stress test across the scalability across large dimensionalities. Again, the accelerator consistently outperforms the GPU: encoder throughput improves by \(1.16\times\)–\(3.66\times\) (peaking at \(D{=}64\mathrm{K}\)), while inference throughput improves by \(1.36\times\)–\(17.51\times\) (peaking at \(D{=}32\mathrm{K}\)). Energy per operation drops by \(13.83\times\)–\(43.7\times\) for encoding and \(16.19\times\)–\(121.87\times\) for inference, with the largest gains occurring at mid-range \(D\). 

\begin{figure}[h]
    \centering
    \includegraphics[width=\linewidth]{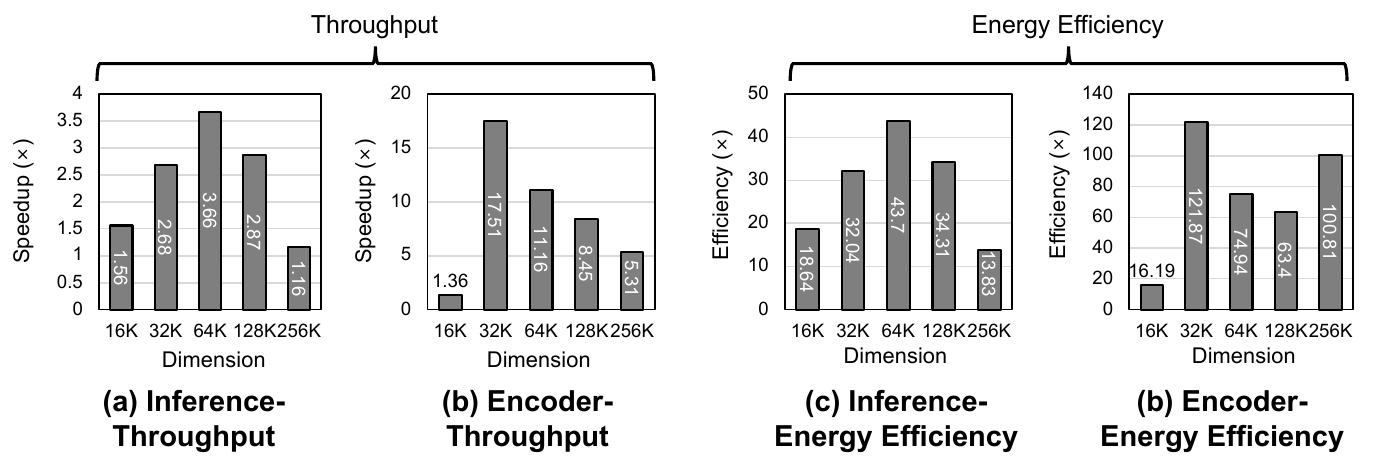}
    \vspace{-2mm}
    \caption{Ablation study comparison of throughput and energy efficiency against an NVIDIA RTX 4090 baseline across large hyperdimensions.}
    \label{fig:hw_ablation}
    \vspace{-3mm}
\end{figure}

Fig.~\ref{fig:hw_ablation} summarizes these trends across various dimensionalities for a complete stress test on any possible complex tasks that need large representational compacity.

\section{Conclusion}

We present the first hardware software co-design accelerated pipeline for GHRR to unify binding, unbinding, and inference, removing the need for expensive permutation logic as well as reducing GHRR's expensive matrix representations to simpler traditional vector representations for inference for exponential reduction during inference time. Additionally, our accelerator achieves over 1.36x–1.56x higher throughput and achieves 16.2x–18.6x better energy efficiency when compared to an NVIDIA RTX 4090 GPU, demonstrating compatibility with custom accelerators. Through this work, we demonstrate that GHRR can be practically deployed as a scalable alternative to traditional HDC architectures through flattening for inference and provides a promising foundation for future, energy-efficient brain-inspired computing architectures on custom or emerging hardware.

\section{Acknowledgment}
This work was supported in part by the DARPA Young Faculty Award, the National Science Foundation (NSF) under Grants \#2127780, \#2319198, \#2321840, \#2312517, and \#2235472, the Semiconductor Research Corporation (SRC), the Office of Naval Research through the Young Investigator Program Award, and Grants \#N00014-21-1-2225 and \#N00014-24-1-2547, Army Research Office Grant \#W911NF2410360. Additionally, support was provided by the Air Force Office of Scientific Research under Award \#FA9550-22-1-0253.

\clearpage
\bibliographystyle{ACM-Reference-Format}
\bibliography{main_bib}

\end{document}